\documentclass[
aps,
superscriptaddress,
reprint,
nofootinbib,
amsmath,amssymb,
physrev,
]{revtex4-2}

\usepackage{graphicx}
\usepackage{color}
\usepackage[utf8]{inputenc}
\usepackage[colorlinks]{hyperref}
\usepackage{bm}% bold math

\usepackage{qcircuit}

\usepackage[linesnumbered, ruled]{algorithm2e}
\usepackage{physics}
\usepackage[version=4]{mhchem}
\usepackage{multirow} 
\usepackage{diagbox}

\newtheorem{thm}{Theorem}

\newcommand{\mr}[1]{\mathrm{#1}}
\newcommand{\mc}[1]{\mathcal{#1}}

\newcommand{\prepare}[0]{\texttt{PREPARE} }
\newcommand{\ry}[0]{R_y}
\newcommand{\rz}[0]{R_z}

\begin{document}
\title{Classical variational optimization of PREPARE circuit \\for quantum phase estimation of quantum chemistry Hamiltonians}

\newcommand{\qunasys}{QunaSys Inc., Aqua Hakusan Building 9F, 1-13-7 Hakusan, Bunkyo, Tokyo 113-0001, Japan}
\newcommand{\handai}{Graduate School of Engineering Science, Osaka University, 1-3 Machikaneyama, Toyonaka, Osaka 560-8531, Japan}
\newcommand{\qiqb}{Center for Quantum Information and Quantum Biology, Osaka University, 1-3 Machikaneyama, Toyonaka, Osaka 560-8531, Japan}
\newcommand{\riken}{Center for Quantum Computing, RIKEN, Wako Saitama 351-0198, Japan.}
\newcommand{\sakigake}{JST, PRESTO, 4-1-8 Honcho, Kawaguchi, Saitama 332-0012, Japan}

% 順番は後で相談
\author{Hayata Morisaki}
\email{hayata.morisaki.osaka@gmail.com}
\affiliation{\qunasys}
\affiliation{\handai}

\author{Kosuke Mitarai}
\affiliation{\handai}
\affiliation{\qiqb}

\author{Keisuke Fujii}
\affiliation{\handai}
\affiliation{\qiqb}
\affiliation{\riken}

\author{Yuya O. Nakagawa}
\email{nakagawa@qunasys.com}
\affiliation{\qunasys}

\begin{abstract}
We propose a method for constructing \texttt{PREPARE} circuits for quantum phase estimation of a molecular Hamiltonian in quantum chemistry by using variational optimization of quantum circuits solely on classical computers.
The \texttt{PREPARE} circuit generates a quantum state which encodes the coefficients of the terms in the Hamiltonian as probability amplitudes and plays a crucial role in the state-of-the-art efficient implementations of quantum phase estimation.
We employ the automatic quantum circuit encoding algorithm [Shirakawa \textit{et al.}, arXiv:2112.14524] to construct \texttt{PREPARE} circuits, which requires classical simulations of quantum circuits of $O(\log N)$ qubits with $N$ being the number of qubits of the Hamiltonian.
The generated \texttt{PREPARE} circuits do not need any ancillary qubit.
We demonstrate our method by investigating the number of $T$-gates of the obtained \texttt{PREPARE} circuits for quantum chemistry Hamiltonians of various molecules, which shows a constant-factor reduction compared to previous approaches that do not use ancillary qubits.
Since the number of available logical qubits and $T$ gates will be limited at the early stage of the fault-tolerant quantum computing, the proposed method is particularly of use for performing the quantum phase estimation with such limited capability.
\end{abstract}

\maketitle

%%%%% -------------------- %%%%%
\section{Introduction \label{sec:intro}}

Quantum phase estimation (QPE)~\cite{kitaev1995quantum, Nielsen2011, cleve1998} is one of the most important algorithms to leverage the computational power of quantum computers. 
It can calculate eigenenergies of an input Hamiltonian $H$ by simulating the dynamics $e^{iHt}$ for time $t$ (or similar unitaries) on quantum computers and is expected to achieve the exponential speedup over classical computers for certain cases.
Especially, quantum phase estimation for the Hamiltonians in quantum chemistry and condensed matter physics~\cite{PhysRevLett.83.5162, Berry2019qubitizationof, lee2021even, PhysRevResearch.3.033055, yoshioka2022hunting,Babbush2018} has been featured because of its possibility to realize the speedup and its industrial importance.

It is considered that QPE for Hamiltonians that are intractable with the current classical computers can only be performed with fault-tolerant quantum computation (FTQC) since complicated quantum circuits are needed. 
One of the most promising and featured schemes to realize FTQC is the surface code, where quantum circuits must be decomposed into Clifford gates and $T$ gates which constitute the universal gate set.
The fault-tolerant execution of $T$ gate needs the magic-state distillation and is the most costly part of FTQC.
Therefore, the number of $T$ gates in a quantum circuit is regarded as an indicator of the quantum computational cost of algorithms.

The first resource analysis of QPE for quantum chemistry calculations \cite{reiher2017elucidating} was based on the Trotterization of the time evolution operator $e^{iHt}$.
It was shown that the number of $T$ gates becomes as large as $10^{15}$ for a Hamiltonian that cannot be solved by the current classical computers.
Recently, the so-called qubitization technique~\cite{berry2018improved, PhysRevLett.121.010501} has been explored to reduce the quantum computational cost for QPE.
The state-of-the-art method can reduce the number of $T$ gates by several orders of magnitudes compared with the Trotterization-based implementation \cite{Babbush2018, Berry2019qubitizationof, lee2021even, PhysRevResearch.3.033055, yoshioka2022hunting}.

In those qubitization-based approaches, the problem Hamiltonian $H$ is block-encoded in a unitary matrix $U$ \cite{Martyn2021}; $U$ is a matrix larger than $H$ and is proportional to $H$ in a certain subspace.
Under an assumption that Hamiltonians in quantum chemistry and condensed matter physics can be expressed as a linear combination of the polynomial number of Pauli operators, the most popular way for its block-encoding is through the linear combination of unitaries (LCU) which uses the so-called \texttt{PREPARE} and \texttt{SELECT} operations~\cite{berry2018improved,Berry2019qubitizationof,Babbush2018,lee2021even}.
\texttt{PREPARE} is an operation that prepares a quantum state with amplitudes proportional to coefficients of Pauli operators in the Hamiltonian (the detailed definition is given later).
\texttt{SELECT} is a sequence of controlled Pauli gates.
Reducing the numbers of gates and ancillary qubits for implementing \prepare and \texttt{SELECT} operations is thus important to realize more efficient QPE.

In this work, we explore how we can implement the \prepare operation more efficiently with respect to the required number of ancillary qubits.
Babbush \textit{et al.}~\cite{Babbush2018} proposed a sophisticated implementation of \texttt{PREPARE} which makes the number of $T$ gates linear in the number of the Pauli operators contained in the Hamiltonians.
It is accomplished by utilizing the so-called quantum read-only memory (QROM) and combining it with alias sampling method \cite{Walker1974}.
However, their method requires a larger number of ancillary qubits than what is naively required for \texttt{PREPARE}.
Although the number of additional ancillary qubits scales logarithmic to the system size and the desired error, for Hamiltonians with system size corresponding to $O(100)$ qubits, it almost doubles the number of qubits in \prepare \cite{lee2021even}. 
There may be a situation where we want to keep the number of qubits as low as possible even with an increased $T$ count, when we consider the early stage of the realization of FTQC (pre-FTQC or early-FTQC era, see e.g. Refs.~\cite{Zhang2022computingground, akahoshi2023partially, kuroiwa2024averaginggateapproximationerror}).
This motivates us to seek the efficient implementation of \prepare without any ancillary qubit.

Here, we propose a scheme to construct \texttt{PREPARE} without any ancillary qubit by performing variational optimization of a quantum circuit on classical simulators.
The framework is scalable because physically motivated Hamiltonians can usually be written as a linear combination of a polynomial number of Pauli operators with respect to its system size and thus the number of qubits involved in \prepare is only logarithmic to its system size.
Specifically, we assume the ansatz quantum circuit for approximating \prepare circuits and employ the automatic quantum circuit encoding (AQCE) algorithm proposed in Ref.~\cite{shirakawa2021automatic}, which systematically optimizes the ansatz circuit to generate a target quantum state.
For numerical illustrations, we apply the proposed scheme to construct \texttt{PREPARE} for quantum chemistry Hamiltonians of various molecules.
We find that our approach can find a circuit with the number of T gates scaling linearly to the number of terms in the Hamiltonian.
More concretely, we can obtain a circuit with about $10^6$ T gates for 20-qubit Hamiltonian.
This corresponds to about a two-fold reduction compared to the naive construction that combines the technique of exactly decomposing \prepare into single-qubit rotation gates and CNOT gates without using ancillary qubits \cite{iten2016quantum}, and the technique of decomposing single-qubit rotation gates into Clifford+$T$ gates \cite{ross2014optimal}.
Our method is practically useful to reduce the quantum computational cost of QPE and paves the way for utilizing the first-generation quantum computers with a limited number of qubits and $T$ gates.

This paper is organized as follows.
In Sec.~\ref{sec:preliminary}, we review the preliminary background of this study.
We define the \prepare operation and explain the AQCE algorithm proposed in Ref.~\cite{shirakawa2021automatic}, which is central to our study.
In Sec.~\ref{sec:proposal}, we explain our proposal to construct \prepare without resorting to any ancillary qubit.
Section~\ref{sec:numerics} shows the numerical illustration of our proposal.
Conclusion and outlook are presented in Sec.~\ref{sec:conclusion}.

%%%%% -------------------- %%%%%
\section{Preliminaries \label{sec:preliminary}} 
In this section, we review several preliminaries to explain our method.
First, we review the definition of \prepare and previous proposals for its implementation.
Second, we review the automatic quantum circuit encoding (AQCE) algorithm~\cite{shirakawa2021automatic} that seeks a quantum circuit to generate a given quantum state.

\subsection{\texttt{PREPARE} circuit}
Let us consider an $n$-qubit Hamiltonian in the form of a liner combination of unitaries,
\begin{align} \label{eq:LCU ham.}
 H = \sum_{l=0}^{L-1} c_l U_l,
\end{align}
where $c_l > 0$ is a coefficient and $U_l$ is a unitary operator on $n$ qubits satisfying $U_l^2=I$.
When the Hamiltonian is written in the sum of $L$ Pauli operators, $H = \sum_{l=0}^{L-1} d_l P_l$, 
where $P_l\in \{I,X,Y,Z\}^{\otimes n}$ is the $n$-qubit Pauli operator, one can always assume that the Hamiltonian takes the form of \eqref{eq:LCU ham.} by setting $c_l = |d_l|$ and $U_l = \mr{sign}(d_l)P_l$. 
A \prepare operation for $\{c_l \}_{l=1}^L$ is defined on $m \geq \lceil \log_2(n) \rceil$ qubits as, 
\begin{align}\label{eq:prepare-definition}
 \texttt{PREPARE}\ket{0}^{\otimes m} = \frac{1}{\sqrt{\sum_l c_l}}\sum_{l=0}^{L-1} \sqrt{c_l}\ket{l},
\end{align}
where $\ket{l}$ is a computational basis state for integer $l$ on $m$ qubits.
\prepare plays a crucial role in performing the LCU protocol~\cite{berry2018improved,Berry2019qubitizationof,Babbush2018,lee2021even} which enables the block-encoding of physical Hamiltonians.

There are various methods to implement \texttt{PREPARE}.
Note that we can only achieve approximate version $\texttt{PREPARE}'$ of \prepare in actual implementations.
Namely, we seek to realize
\begin{equation}
 \prepare' \ket{0}^{\otimes m} = \sum_{l=0}^{L-1} e_l \ket{l}, \: \sum_{l=0}^{L-1} |e_l|^2 = 1
\end{equation}
where $e_l$ is tuned as close to $\sqrt{c_l} / \sqrt{\sum_l c_l}$ as possible.
We interpret this as an approximation of \prepare by defining $c_l' = |e_l|^2 (\sum_l c_l)$, namely,
\begin{align} \label{eq:prepare_approx}
    \prepare' \ket{0}^{\otimes m} = \frac{1}{\sqrt{\sum_l c_l}} \sum_{l=0}^{L-1} \sqrt{c_l'}\ket{l}.
\end{align}
Note that $c_l' \geq 0$ and $\sum_l c_l' = \sum_l c_l$ hold.
We say that $\prepare'$ approximates the original \prepare within the error $\epsilon \ge 0$ when
\begin{equation} \label{eq:error def}
 \max_{l} \abs{c_l -c_l'} \leq \epsilon.
\end{equation}

One of the simplest ways to implement \prepare without requiring any ancillary qubit is to apply controlled $R_y$ gates sequentially, whose angles are determined by the binary partition of the classical data, $c_0, \cdots, c_{L-1}$.
This method has been explicitly considered in Ref.~\cite{Mottonen2005Transformation, Bergholm2005, iten2016quantum}.
The number of $T$ gates required in this implementation is $O\left(L\log(L) + L\log(1/\epsilon)\right)$, where $\epsilon$ is the approximation error of \prepare (see Eq.~\eqref{eq:error def}).
While this approach provides a procedure to construct a circuit to generate an arbitrary state, there can be room for problem-dependent optimization.
Another strategy for implementing \prepare is to use quantum read-only memory (QROM)~\cite{Babbush2018}, which loads information of $\{ c_l \}_{l=0}^{L-1}$ in ancillary qubits. 
Babbush \textit{et al.}~\cite{Babbush2018} proposed an efficient implementation of \prepare for general $\{ c_l \}_{l=0}^{L-1}$ based on QROM, which requires $O(\log L + \log(1/\epsilon))$ ancillary qubits and $O(L + \log(1/\epsilon))$ $T$ gates (see Appendix~\ref{appsec:qrom based prepare}).
It should be noted that several improvements on the methods based on QROM for specific types of Hamilonians, like quantum chemistry Hamiltonians, are discussed in Refs.~\cite{Berry2019qubitizationof, lee2021even, yoshioka2022hunting}.

In this study, we aim at implementing \prepare circuits without any ancillary qubit with the use of the variational optimization of quantum circuits on classical computers, based on the AQCE algorithm~\cite{shirakawa2021automatic} explained in the following.

\subsection{Automatic quantum circuit encoding (AQCE) algorithm \label{subsec:AQCE}}

The AQCE algorithm~\cite{shirakawa2021automatic} is a method to construct a $m$-qubit quantum circuit $\mathcal{C}$ which satisfies $\mathcal{C}\ket{0}\approx \ket{\Psi}$ for a given $m$-qubit quantum state $\ket{\Psi}$.
Here, we review the algorithm for the case where the circuit $\mathcal{C}$ is fully comprised of two-qubit gates, and classical computers are used to run the whole algorithm.
We note that a similar algorithm to find an approximate circuit $\mc{C}$ for a given state $\ket{\Psi}$ is proposed in Ref.~\cite{lin2021real}.

The AQCE constructs the circuit $\mathcal{C}$ as a sequence of $M$ two-qubit gates $\mc{U}_s$ ($s=1,\cdots,M$):
\begin{equation}
 \mathcal{C} = \mc{U}_M \cdots \mc{U}_{2} \mc{U}_1. 
\end{equation}
All the components of the two-qubit gate $\mc{U}_s$ ($4 \times 4$ matrix, generally with complex-valued entries) as well as the qubit indices on which the gate acts are optimized iteratively.
The optimization of $\mc{U}_s$ for a fixed $s$ is performed by maximizing the fidelity $F(\mc{U}_s) = |\mel{\Psi}{\mc{C}}{0}| = |\mel{\Psi}{\mc{U}_M \cdots \mc{U}_{s+1} \cdot \mc{U}_s \cdot \mc{U}_{s-1} \cdots \mc{U}_1}{0}|$ while keeping the other $\mc{U}_{s'\neq s}$ fixed.
When we assume $\mc{U}_s$ acts on qubits $i_s$ and $j_s$, the maximal value of $F(\mc{U}_s)$ can be achieved by performing the singular value decomposition of $4 \times 4$ matrix $\rho_{s,i_s,j_s}$ defined as,
\begin{align}
\rho_{s,i_s,j_s} &= \Tr_{ \{1,2,\cdots,m\} \setminus \{ i_s, j_s\} }  \left( \ket{\Phi_s} \bra{\Psi_s} \right),
\end{align}
where $\ket{\Psi_{s}} = \mc{U}_{s+1}^\dag \cdots \mc{U}_M^\dag \ket{\Psi}$ and $ \ket{\Phi_{s}} = \mc{U}_{s-1} \cdots \mc{U}_1 \ket{0}$ \cite{shirakawa2021automatic}. 
Namely, when the singular value decomposition of $\rho_{s,i_s,j_s}$ is $\rho_{s,i_s,j_s} = XDY$, where $X$ and $Y$ are unitary matrices and $D$ is a diagonal matrix with non-negative entries, choosing $\mc{U}_s = Y^\dag X^\dag$ achieves the maximum of $F(\mc{U}_s)$.
The AQCE algorithm optimizes each $\mc{U}_s$ by trying all possible pairs of the qubits $(i_s, j_s)$ and choosing the pair that gives the largest fidelity.
This completes the optimization of the gate $\mc{U}_s$ including the choice of qubit indices.
We call this procedure to optimize $\mc{U}_s$ the subroutine $\texttt{Optimize}(\mc{U}_s)$.

The whole AQCE algorithm is as follows:
\begin{enumerate}
 \item We add $M_0$ two qubit gates to the circuit $\mc{C}$ as $\mc{C} = \mc{U}_{M_0} \cdots \mc{U}_1$ with the initialization $\mc{U}_1 = \cdots = \mc{U}_{M_0} = I$ (identity operator) and set $M \gets M_0$.
 \item We optimize the gates $\mc{U}_s$ sequentially from $s=1$ to $s=M$ by calling the subroutine $\texttt{Optimize}(\mc{U}_s)$. The sequential optimization from $s=1$ to $s=M$ is called forward update. After the forward update, we optimize the gates $\mc{U}_s$ sequentially from $s=M$ to $s=1$, which is called backward update. We repeat the forward and backward updates, which is called \textit{sweep}, $N$ times.
 \item We check if $\mc{C}\ket{0}$ approximates the target state $\ket{\Psi}$ within the desired accuracy $\epsilon$ by using some distance metric $d(\ket{\Psi}, \mc{C}\ket{0})$.
 If the accuracy is met, or $d(\ket{\Psi}, \mc{C}\ket{0}) \leq \epsilon$, the algorithm stops and outputs the circuit $\mc{C}$.
 If this is not the case, we add new $\delta M$ gates to the circuit $\mc{C}$ as $\mc{C}' = \mc{U}_{M + \delta M} \cdots \mc{U}_{M + 1} \cdot \mc{C}$ with the initialization $\mc{U}_{M+1} = \cdots = \mc{U}_{M+\delta M} = I$. We call $\texttt{Optimize}(\mc{U}_s)$ sequentially for $s=M+1, \cdots, M+\delta M$.
 \item We set $M \gets M + \delta M$ and repeat steps 2 and 3.
\end{enumerate}

In the original proposal of AQCE~\cite{shirakawa2021automatic}, it was numerically shown that this algorithm can create quantum circuits for representing the ground states of several Hamiltonians in physics and a quantum state which encodes some classical data (image data).
In the following section, we propose to utilize AQCE in constructing \prepare circuits for quantum chemistry Hamiltonians.

%%%%% -------------------- %%%%%
\section{Our proposal \label{sec:proposal}}
\begin{figure}[t]
\includegraphics[width=\linewidth]{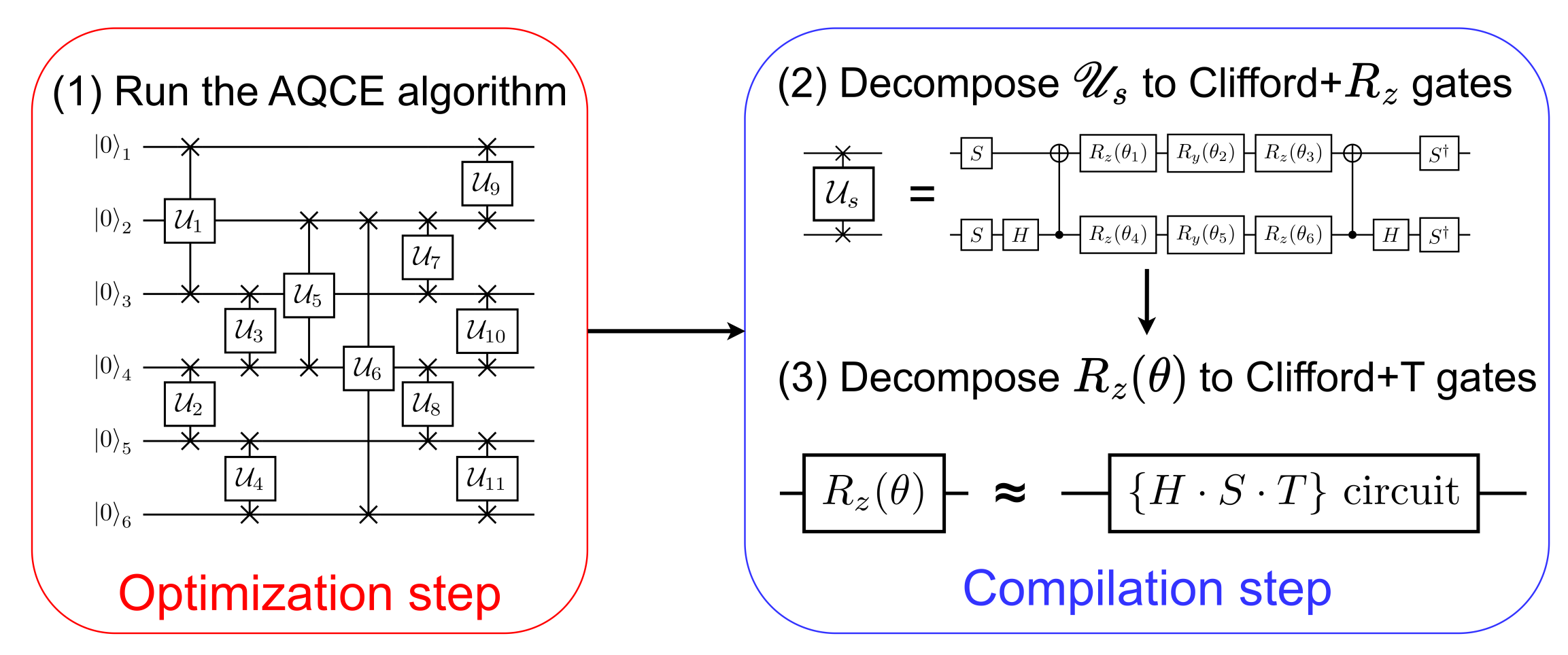}
\caption{Schematic picture of our algorithm to construct \prepare circuit without any ancillary qubit and its Clifford+$T$ decomposition.
(1) We first run AQCE for the target quantum state $\ket{\Psi}$.
(2) Two-qubit gates in the resulting circuit of AQCE are then decomposed into six one-qubit rotation gates and Clifford gates.
(3) Finally, the one-qubit rotation gates are decomposed into Clifford+$T$ gates by using the algorithm proposed in Ref.~\cite{ross2014optimal}.
\label{fig:schematics}}
\end{figure}

In this section, we propose a method to construct $\texttt{PREPARE}$ circuit and its decomposition into Clifford gates and $T$ gates (i.e., Clifford+$T$ gate decomposition) for a given Hamiltonian.
Our method leverages the AQCE algorithm reviewed in the previous section and combines it with the Clifford+$T$ gate decomposition of a single-qubit rotation gate~\cite{ross2014optimal}.

Our algorithm to obtain \prepare for the Hamiltonian~\eqref{eq:LCU ham.} with an error $\epsilon$ is as follows (see also Fig.~\ref{fig:schematics}).
\begin{enumerate}
 \item Run the AQCE algorithm with the target state,
 \begin{equation}
      \ket{\Psi} = \frac{1}{\sqrt{\sum_l c_l}}\sum_{l=0}^{L-1} \sqrt{c_l}\ket{l},
 \end{equation}
 and obtain the circuit $\mc{C}$.
 The distance metric is chosen as $d(\ket{\Psi}, \mc{C}\ket{0}) = \max_l {\abs{c_l - c_l'}}$ where we denote  $\mc{C}\ket{0} = \frac{1}{\sqrt{\sum_{l} c_l}} \sum_{l=0}^{L-1} \sqrt{c_l'} \ket{l}$.
 Since we later decompose the resulting circuit $\mc{C}$ into a Clifford+$T$ gate set, which introduces some error, the AQCE algorithm should be run with slightly stronger convergence criteria $d(\ket{\Psi}, \mc{C}\ket{0}) < \epsilon'\leq \epsilon$ when we seek to find a final circuit satisfying $d(\ket{\Psi}, \mc{C}\ket{0}) \leq\epsilon$.
 Exploiting the fact that the target state $\ket{\Psi}$ is real in our case, we take real $\mc{U}_s$. Note that the singular value decomposition of real matrices can be performed by real $X$, $Y$, and $D$.
 \item Decompose each real two-qubit gates $\mc{U}_s$ for $s=1,\cdots,M$ into Clifford gates and six $R_z$ gates as Fig.~\ref{fig:schematics} (2) with the method developed in Ref.~\cite{vatan2004optimal} (see also Appendix~\ref{appsec:O4 decompose}).
 \item Decompose each of six $R_z$ gates approximately into the Hadamard, $S$ and $T$ gates by the Ross-Selinger's algorithm~\cite{ross2014optimal}.
 The algorithm can generate an approximate circuit $U_\mr{approx}$ which satisfies $\| R_z - U_\mr{approx}\| \leq \epsilon_T$ for a given $R_z$ gate where $\|\cdot\|$ is the operator norm.
 To make the resulting circuit satisfy $\max_l | c_l-c_l'| \leq \epsilon$, we need to use sufficiently small $\epsilon_T$.
 To find optimal $\epsilon_T$, we use the same $\epsilon_T$ for decomposing all $R_z$ gates in the circuit and perform the binary search for the largest $\epsilon_T$ that satisfies the condition above.
\end{enumerate}
We note that the whole procedure is assumed to be performed on classical computers and its feasibility for large quantum systems of interest is discussed later.

The above algorithm is efficient on classical computers with respect to the system size of the target Hamiltonian.
This is because the number of qubits $m$ required for \prepare circuits is $m=\lceil\log_2(L)\rceil$, and $L$ scales typically as $O(\mr{poly}(n))$ for $n$-qubit systems in physically or chemically interesting Hamiltonians.
We therefore have $m = O(\log n)$, which means that
the classical computational cost for simulating $m$-qubit quantum circuits in our algorithm is polynomial in $n$.
We also present the actual number of qubits $m$ in Fig.~\ref{fig:number of qubits required} for concrete molecular systems in the following section.

The hyperparameter $\epsilon'$ determines the balance between the increase of $T$ gates arising from the intrinsic number of two-qubit gates used in the AQCE algorithm and that from the decomposition of $R_z$ gates.
The number of $T$ gates required in our algorithm can roughly be written as $6M n_T$, where $M$ is the total number of two-qubit gates in AQCE and $n_T$ is the typical number of $T$ gates used to decompose one $R_z$ gate.
Smaller $\epsilon'$ would result in a larger $M$ and relaxed $\epsilon_T$ (and thus smaller $n_T$), and vice versa.
Although it is possible to optimize $\epsilon'$ to achieve a minimal number of $T$ gates, we leave such an approach as a possible future work.
There are two main difficulties in building such a strategy.
First, since AQCE is a variational and heuristic method, it is generally difficult to predict $M$ for a given $\epsilon'$.
Second, the Ross-Selinger decomposition uses the error defined by the operator norm, which cannot be straight-forwardly converted into the metric $\max_l | c_l-c_l'|$ that is relevant to \prepare.

These considerations together make it difficult to provide a rigorous analytical statement about the error-scaling (or the required number of $T$ gates) of our proposal.
We, therefore, conduct numerical experiments in the next section to demonstrate that it can indeed generate circuits with less number of gates than the naive method that combines \cite{iten2016quantum} and \cite{ross2014optimal}.

%%%%% -------------------- %%%%%
\section{Numerical calculation \label{sec:numerics}}
In this section, we present a numerical illustration of our algorithm for quantum chemistry Hamiltonians of small molecules, including hydrogen chain molecules which are benchmark systems in computational quantum chemistry. 
We first investigate the number of qubits $m$ for large quantum systems that are beyond the capability of classical computers and find that $m$ is still feasible even if the systems itself is large enough.
We then numerically show that our method can construct \prepare circuits whose number of $T$ gates is much smaller than the naive implementation without ancillary qubits.

We consider hydrogen chain molecules, \ce{H2}, \ce{H4}, \ce{H6}, \ce{H8},  \ce{H10}, \ce{H12}, and \ce{H14}, with the inter-atomic distance being 1\AA.
We also consider \ce{LiH}, \ce{H2O}, \ce{NH3}, \ce{CH4}, \ce{CO}, \ce{H2S}, \ce{C2H2} (acetylene), \ce{CO2}, and \ce{C6H6} (benzene) molecules whose geometries are listed in Table~\ref{tab: geometries}.
We perform the Hartree-Fock calculation with the STO-3G basis set to construct chemistry Hamiltonians and employ Jordan-Wigner transformation~\cite{Jordan1928} to map the chemistry Hamiltonian into the qubit Hamiltonian.
The number of qubits $n$ of the resulting Hamiltonian for $\ce{H}_k$ molecule is $2k$ and those for the other molecules are listed in Table~\ref{tab: geometries}.
Further details of the numerical calculation are described in Appendix~\ref{appsec:details of numerics}.
In addition, we consider the nitrogenase FeMo cofactor (FeMoco) by using the data of the Hamiltonian presented in Li~\textit{et al.}~\cite{Li2019Femoco}.
FeMoco is considered one of the most important applications of quantum computers in industry, and its Hamiltonian studied in Ref.~\cite{Li2019Femoco} is described by 152 qubits.

\begin{figure}
 \includegraphics[width=0.49\textwidth]{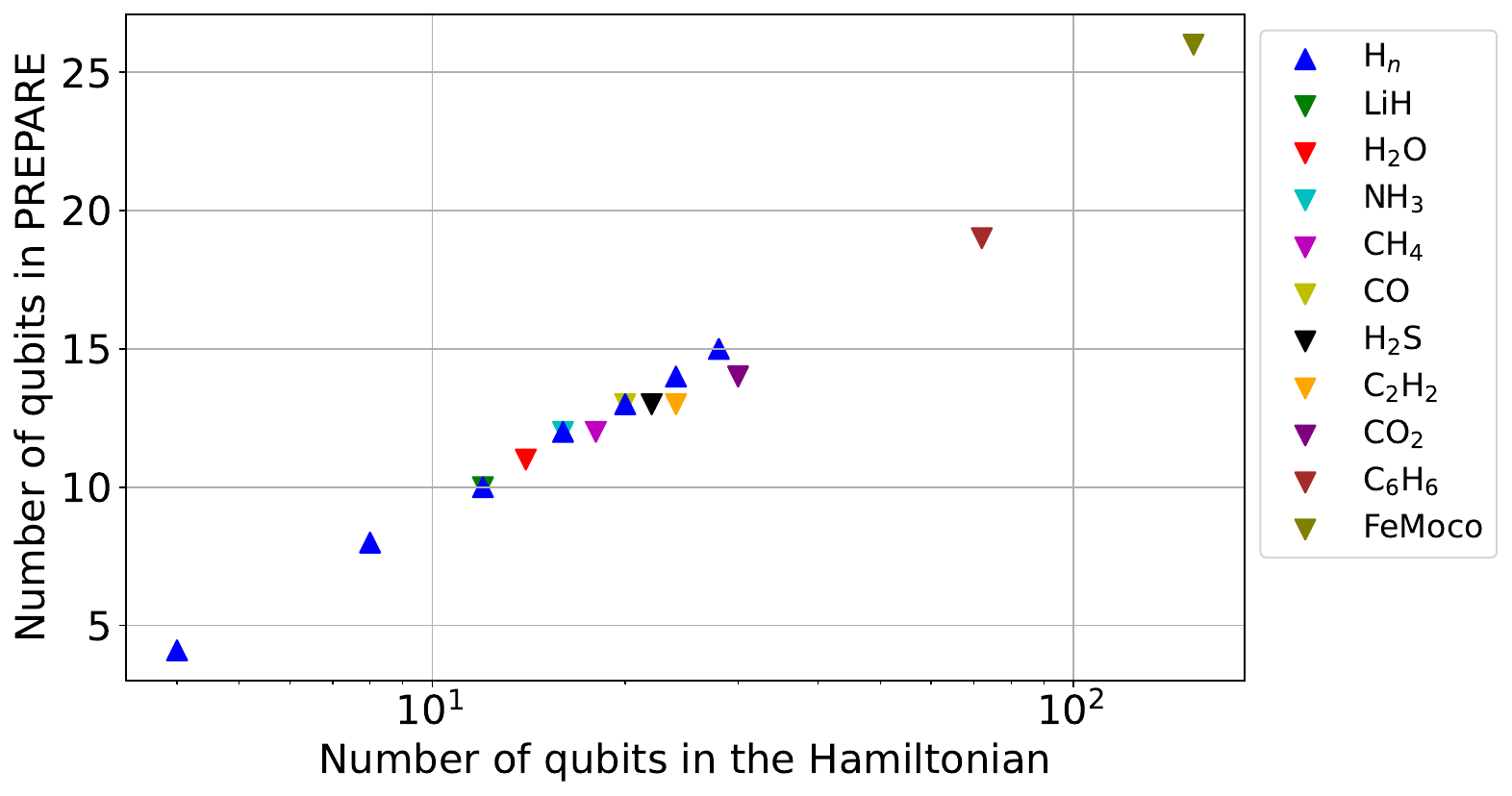}
 \caption{Number of qubits $m = \lceil \log_2 L \rceil$ required to construct \prepare circuit for each molecule.
 \label{fig:number of qubits required}}
\end{figure}

First, we evaluate the number of qubits $m$ for \prepare circuits by examining the number of Pauli operators $L$ in the Hamiltonian.
The required number of qubits $m = \lceil \log_2(L) \rceil$ for \prepare circuits is shown in Fig.~\ref{fig:number of qubits required}.
It can be seen that $m$ increases only with $O(\log(n))$ as expected.
Moreover, even when the system is large enough and classical exact diagonalization is not easy ($n\gtrsim 28$), $m$ is still around 15 to 26.
The classical simulation of quantum circuits of this scale is feasible.
This illustrates the possibility of utilizing our method in practical applications of quantum phase estimation to classically intractable large molecular systems.

\begin{table*}[t]
    \centering
    \begin{tabular}{c|c||c|c|c|c|c|c|c|c|c|c|c|c}
        \hline
        \multicolumn{2}{c||}{Molecule} & \ce{H2} & \ce{H4} & \ce{H6} & \ce{H8} & \ce{H10} & \ce{LiH} & \ce{H2O} & \ce{NH3} & \ce{CH4} & \ce{CO} & \ce{H2S} & \ce{C2H2} \\
        \hline \hline
        \multicolumn{2}{c||}{\#terms $L$} & 14 & 184 & 918 & 2912 & 7150 & 630 & 1085 & 3056 & 2211 & 4426 & 6245 & 5184 \\
        \hline \hline
        \multirow{3}{*}{\#T gates} 
        & Our method & 1672 & 27628 & 149284 & 666510 & 1307480 & 144784 & 345906 & 745018 & 673784 & 1426300 & 1444732 & 1429838 \\ 
        \cline{2-14}
        & Naive method & 1256 & 58844 & 272908 & 1339579 & 2735622 & 288865 & 606764 & 1315655 & 1301066 & 2751921 & 2726869 & 2439315\\
        \cline{2-14}
        & QROM method & 650 & 1942 & 5282 & 13594 & 30831 & 3992 & 6125 & 14254 & 10730 & 19905 & 27341 & 22895 \\
        \hline \hline
        \multirow{3}{*}{\begin{tabular}{c} \#ancillary\\ qubits \end{tabular}} 
        & Our method   \\
        \cline{2-14}
        & Naive method \\
        \cline{2-14}
        & QROM method & 47 & 55 & 59 & 63 & 66 & 57 & 64 & 65 & 63 & 68 & 70 & 66\\
        \hline
    \end{tabular}
    \caption{Number of $T$ gates and ancillary qubits for each molecule in the three methods.
    Note that our method and the naive method are both ancilla-free constructions.}
    \label{tab:T_count_ancilla}
\end{table*}

\begin{figure*}[t] \includegraphics[width=0.45\textwidth]{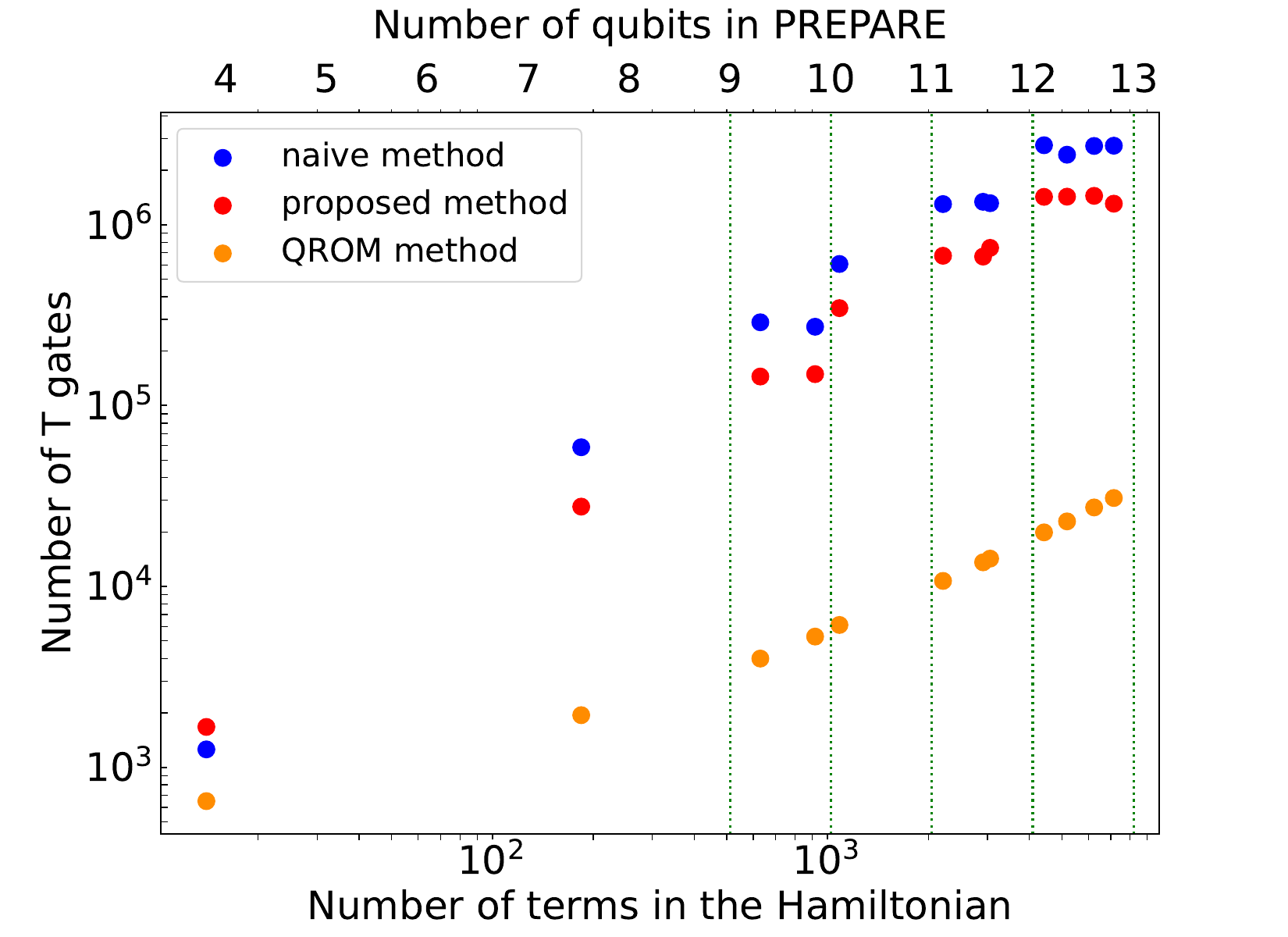}
\includegraphics[width=0.45\textwidth]{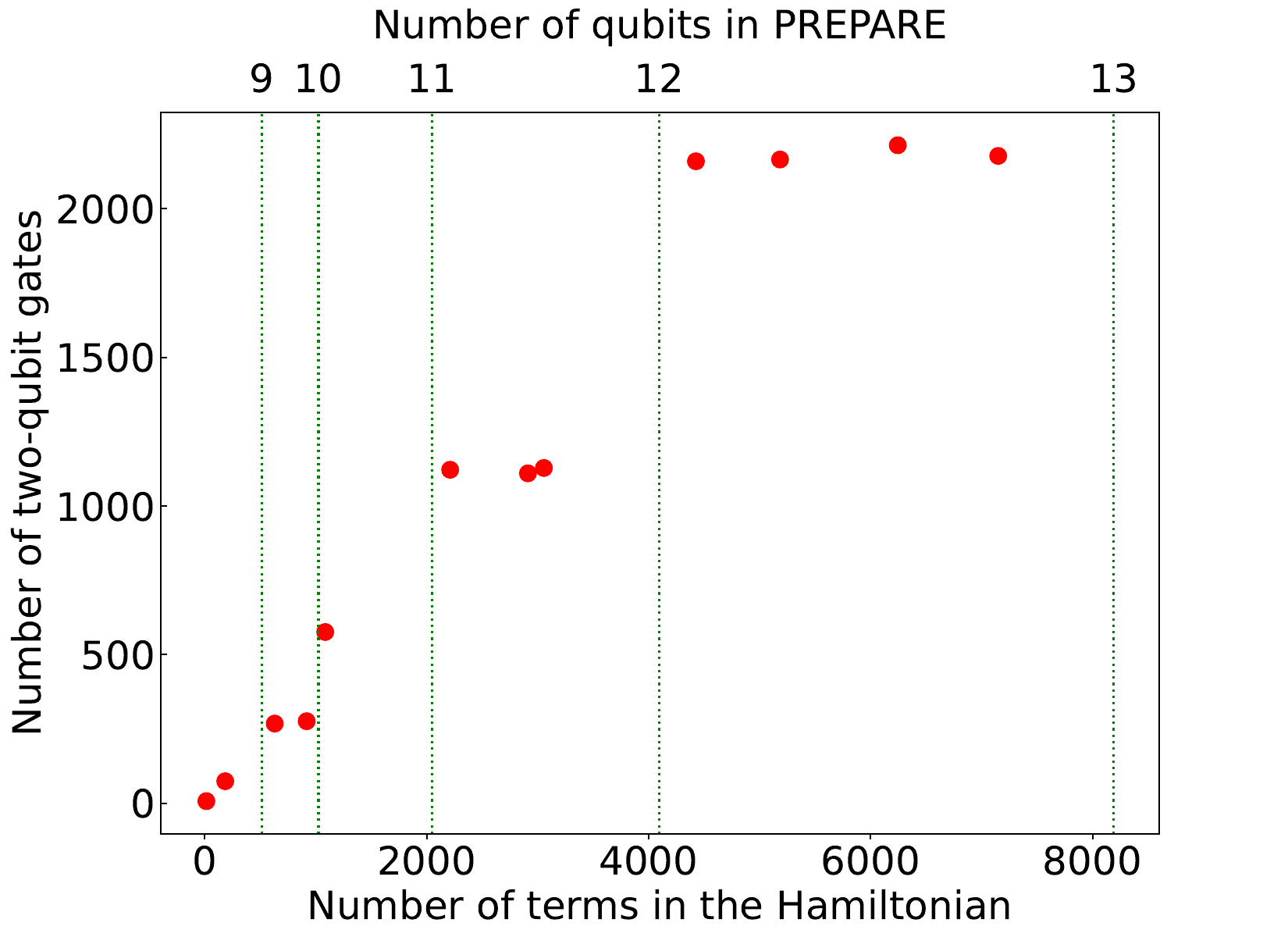}
  \caption{\label{fig:T_count} (a) Number of $T$ gates of \prepare circuits for the molecules shown in Tab.~\ref{tab:T_count_ancilla} as a function of the number of terms in the Hamiltonian $L$.
  (b) Number of two-qubit gates $\mc{U}_s$ in the quantum circuit generated by AQCE in our algorithm.}
\end{figure*}

Next, we count the number of $T$ gates and two-qubit gates required in constructing \prepare circuits by our method and compare it with other methods.
The precision $\epsilon$ of approximate \prepare circuits is determined as follows.
Reference~\cite{Babbush2018} shows that the energy difference of two Hamiltonians $H=\sum_l c_l P_l$ and $H'=\sum_l c'_l P_l$ is within $\Delta E$ if
\begin{equation}\label{eq:delta}
 \max_l |c_l - c'_l|  \leq \frac{\sqrt{2} \Delta E}{4L\left(1+\frac{\Delta E^2}{8\lambda^2}\right)},
\end{equation}
where $\lambda = \sum_l |c_l|$.
We put $\Delta E = 0.0016$ Hartree, which is called chemical accuracy in quantum chemistry calculations, on the right-hand side of the above equation to determine the desired error $\epsilon$ in approximating \texttt{PREPARE}.
We sort $\{ c_l \}_{l=0}^{L-1}$ in descending order before running AQCE since we found it gave higher precision.
The parameters of AQCE in our method are chosen as $(M_0, \delta M, N) = (1,1,100)$ for \ce{H2}, \ce{H4}, \ce{LiH} and $(M_0, \delta M, N) = (12,6,100)$ for other molecules.
The precision for AQCE, or $\epsilon'$ in Sec.~\ref{sec:proposal}, is set to $\epsilon$.
This setting results in a quantum circuit with a slightly better precision than $\epsilon$.
We can then obtain an approximate \prepare circuit with precision $\epsilon$ in the Clifford+$T$ gate set by taking $\epsilon_T$ small enough. 
More concretely, we use binary search to find the largest $\epsilon_T$ that makes the total precision gets smaller than $\epsilon$.
We count the number of $T$ gates only for  \ce{H2}, \ce{H4}, \ce{H6}, \ce{H8}, \ce{H10}, \ce{LiH}, \ce{H2O}, \ce{NH3}, \ce{CH4}, \ce{CO}, \ce{H2S} and \ce{C2H2} molecules that require $m \lesssim 13$ qubits for \prepare circuits, where AQCE converges within the reasonable amount of classical computational time on a single work station.

We compare our method with two methods.
One method is the naive construction obtained through the following steps:
\begin{enumerate}
    \item Exactly decompose \prepare into single-qubits rotation gates and CNOT gates without ancillary qubits by Ref. \cite{iten2016quantum}, which is the state-of-the-art algorithm of this kind.
    \item Decompose each of the single-qubits gates approximately into the Hadamard, $S$ and $T$ gates by the Ross-Selinger's algorithm \cite{ross2014optimal}.
    The precision for decomposing each gate, $\epsilon_T$, is determined with the binary search to satisfy Eq.~\eqref{eq:delta} in the same manner as our algorithm.
\end{enumerate}

The second method which we compare with our method is the QROM-based one proposed in Ref.~\cite{Babbush2018}.
This method requires $m+2\mu+1$ ancillary qubits and
\begin{equation}
4L + 4\mu + 11m - 12 + 2g_T(\epsilon)
\end{equation}
$T$ gates, where $\mu$ determines the precision of \prepare circuit and $g_T(\epsilon)$ is the required number of $T$ gates to decompose some $R_z$ gates in the construction with the precision $\epsilon$.
The details are described in Appendix~\ref{appsec:qrom based prepare}.
Following Ref.~\cite{Babbush2018}, we choose $\mu$ as
\begin{equation}
     \mu = \left\lceil \log_2\left(\frac{\sqrt{2} \Delta E^2}{4\lambda(1+\frac{\Delta E^2}{8\lambda^2})}\right) \right\rceil
\end{equation}
so that the resulting \prepare circuits satisfies Eq.~\eqref{eq:delta}.

TABLE~\ref{tab:T_count_ancilla} and Fig. ~\ref{fig:T_count}(a) show the results of the number of $T$ gates and ancillary qubits for all methods.
It is apparent that our method achieves about a two-fold reduction in the number of $T$ gates than the naive method except for the leftmost point (\ce{H2}).
The QROM-based method \cite{Babbush2018} outperforms our method and the naive one.
This is because it sophisticatedly reduces the $T$ count by the use of ancillary qubits.
However, the number of ancillary qubits required in this method can be large compared to $m$, the minimum number of qubits necessary for \prepare.
For example, 66 ancillary qubits are required to realize \prepare circuits for \ce{H10} molecules (the rightmost point in Fig.~\ref{fig:T_count}).
This number might be problematic for the early-FTQC era.

Next, we also present the number of two-qubit gates, i.e., $M$ in AQCE, in Fig.~\ref{fig:T_count}(b).
This metric can be of interest when we consider the scheme that is out of standard surface-code-based FTQC such as surface codes without magic state distillation \cite{akahoshi2023partially}.
The number of $T$ gates is almost proportional to the number of two-qubit gates generated in AQCE.
It is also seen that the number of two-qubit gates remains almost the same as long as the number of qubits, i.e., $m$ in \prepare does not change.
The number of two-qubit gates is around 2000 for $m=13$, which implies that it might be possible to perform simple demonstrations of the block-encoding for the $\ce{H10}$ molecule when we have two-qubit gates with $\sim 10^{-4}$ error rates assuming all-to-all connection between qubits.
It might be possible on, e.g., near-term ion trap devices or early-FTQC devices employed with $10^4$ physical qubits \cite{akahoshi2023partially}.

\begin{figure}[t]
 \includegraphics[width=0.49\textwidth]{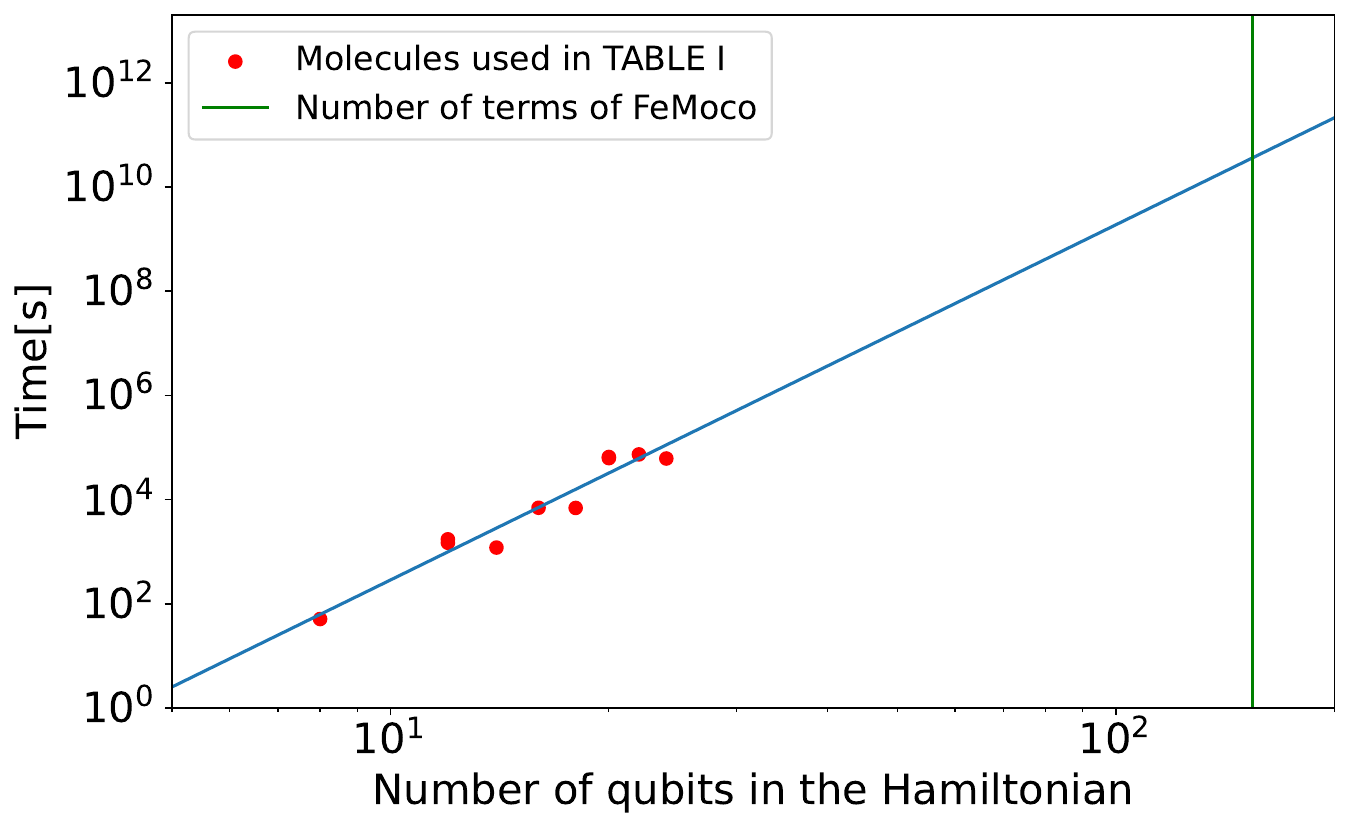}
 \caption{Computation time taken for our method.
The blue line represents the power regression of the red points (data) and is described by the equation $y = 4.37\times 10^{-5} \times x^{6.82}$.}
 \label{fig:time for our method}
\end{figure}

Finally, we discuss the classical computation time for our method. 
Fig.~\ref{fig:time for our method} shows the actual time taken for obtaining \prepare by our method, presented in Fig.~\ref{fig:T_count}.
The CPU used in our calculations is Intel Xeon Gold 5215 and we performed parallel updates of two qubit gates $\mc{U}_s$ up to 48 parallel threads.
As expected, the execution time of our method increases polynomially with the number of qubits in the Hamiltonian.
Nevertheless, the execution time for $\ce{H10}$ (the rightmost red point) already took $6\times 10^5$ seconds (about one day), and it is expected that executing time for FeMoco will require $3\times10^{10}$ seconds (about 1000 years) in a naive extrapolation.
To execute AQCE for large molecular systems such as FeMoco, we need, for example, the parallelization of the AQCE algorithm using large-scale computers and the improvement of optimization strategies.
% We point out that the AQCE algorithm allows parallel processing in several parts other than   optimization of all pairs of the qubits.

%%%%% -------------------- %%%%%
\section{Conclusion \label{sec:conclusion}}
In this study, we propose a method for constructing \prepare circuits without ancillary qubits utilizing the variational algorithms run on classical simulators.
The classical simulation is feasible since the number of qubits of \prepare circuits is $m=O(\log N)$, where $N$ is the system size of the problem Hamiltonian.
More specifically, we employ AQCE for variationally constructing \prepare and the Ross-Selinger algorithm to decompose it to Clifford+$T$ circuits.
To validate our approach, we perform numerical calculations to estimate the number of $T$ gates required for \prepare circuits for quantum chemistry Hamiltonians of various molecules.
We observe the reduction of the number of $T$ gates compared to the naive method \cite{iten2016quantum}.

As future work, it is interesting to apply our method to Hamiltonians other than quantum chemistry Hamiltonians.
For example, the Hamiltonian with many random interaction terms such as Sachdev–Ye–Kitaev model is one candidate.
Another intriguing direction is to seek more efficient protocols to variationally optimize circuits, which is the most time-consuming part of our algorithm.
We use AQCE in this work, but it is certainly not a unique choice.
For example, constructing the circuit by assuming the matrix product state structure might result in faster optimization convergence due to its existing optimization techniques.
Even using AQCE, the protocol can be optimized more.
For example, in the numerical simulation in Sec.~\ref{sec:numerics}, we sort the coefficients $\{ c_l \}_{l=0}^{L-1}$ in descending order based on empirical observation, but other pre-processing may perform better in reducing the number of two-qubit gates and hence the number of steps before the convergence.
It is also notable that we use rather simple method for converting the circuit found by the AQCE algorithm into CNOT and single-qubit gates by converting each two-qubit gate via \cite{vatan2004optimal}. As the number of single-qubit rotations is directly converted to the T count, T count can be further reduced by first performing the compilation of the circuit in such a way that the number of single-qubit rotations is minimized.
Finally, $\texttt{SELECT}$ operation that is not considered in this work should also be optimized for ancilla-free optimization to realize block-encodings in early-FTQC regimes.

\begin{acknowledgements}
This work is supported by MEXT Quantum Leap Flagship Program (MEXT Q-LEAP) Grant No. JPMXS0118067394 and JPMXS0120319794, and
JST COI-NEXT Grant No. JPMJPF2014.
KM is supported by  JST
PRESTO Grant No. JPMJPR2019 and JSPS KAKENHI Grant No. 23H03819.
\end{acknowledgements}

%%%%% -------------------- %%%%%
% \clearpage
\appendix

%%%%% -------------------- %%%%%
\section{Construction of \prepare circuit by Babbush \textit{et al.} \label{appsec:qrom based prepare}}
In this section, we describe a method to construct \prepare circuits based on QROM~\cite{Babbush2018}.
Note that it is referred to as \texttt{SUBPREPARE} in \cite{Babbush2018}.
Although more efficient constructions of \prepare circuits tailored for quantum chemistry Hamiltonian were proposed in Refs.~\cite{Babbush2018,Berry2019qubitizationof},
this construction is generally applicable to any data $\{c_l \}_{l=0}^{L-1}$.

The goal of QROM-based \prepare for the data $\{c_l \}_{l=0}^{L-1}$ is to create a quantum circuit
\begin{eqnarray}
    \texttt{U} \ket{0}^{\otimes m}\ket{0} ^{\otimes N_a} = \frac{1}{\sqrt{\sum_l c_l}} \sum_{l=0}^{L-1} \sqrt{c_l} \ket{l}\ket{\textrm{garbage}_l}, 
\end{eqnarray}
where $N_a$ is the number of ancillary qubits and $\ket{\mr{garbage}_l}$ is some state on $N_a$ qubits.
The presence of the garbage state depending on $l$ does not affect the execution of the block-encoding.
We call the first $m$ qubits in the above equation the ``original" $m$ qubits in the following.
QROM for data $\{ d_l \}_{l=0}^{L-1}$ is a quantum circuit that implements the following transformation: 
\begin{equation}
    \textrm{QROM} \sum_{l=0}^{L-1} \alpha_l \ket{l} \ket{0}^{\otimes \mu} = \sum_{l=0}^{L-1} \alpha_l \ket{l} \ket{d_l},
\end{equation}
where $\alpha_l \in \mathbb{C}$ is an arbitrary coefficient, $\mu$ is the number of qubits for the data register and thereby determines the precision of the loaded data, and $\ket{d_l}$ is the $\mu$-bit representation of $d_l$.
QROM can be realized by additional $m$ work qubits, and the number of $T$ gates is $4L-4$~\cite{Babbush2018}.
We mean by ``work qubits" that those qubits can be released after the operation.

\begin{figure}
  \centering
  \includegraphics[width=\linewidth]{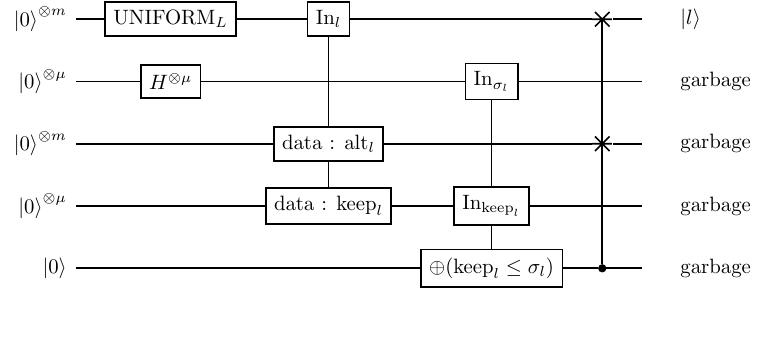}
  \caption{\label{fig:QROM_PREPARE} QROM-based construction of \prepare circuits~\cite{Babbush2018}.}
\end{figure}

The construction of \prepare based on QROM is presented in Fig.~\ref{fig:QROM_PREPARE}.
Apart from the original $m$ qubits for \prepare, it requires $N_a = 2\mu + m + 1$ ancillary qubits and $\max \{m+1, \mu \}$ work qubits.
First, we prepare the uniform superposition of $\ket{l}$ for the original $m$ qubits,
\begin{align}
\begin{split}
    &\textrm{UNIFORM}_L \ket{0}^{\otimes m} \ket{0}^{\otimes (\mu+m+\mu+1)} \\
    &= \sum_{l=0}^{L-1} \frac{1}{L} \ket{l} \ket{0}^{\otimes (\mu+m+\mu+1)}.
\end{split}
\end{align}
This circuit requires $m+1$ work qubits and $4m-4 + 2g_T(\epsilon_z)$ $T$ gates, as shown in Fig.~12 of Ref.~\cite{Babbush2018}.
Here, $g_T(\epsilon_z)$ is the number of $T$ gates required to decompose $R_z(\arccos \frac{2^m-L}{L})$ gates appearing in the construction with the precision
$\epsilon_z$.
We take $\epsilon_z$ as the same as the right-hand side of Eq.~\eqref{eq:delta}.
Next, we use QROM to load the data called $\mr{alt}_l$ ($m$-bit integers) and $\mr{keep}_l$ ($\mu$-bit integers).
The values of $\mr{alt}_l$ and $\mr{keep}_l$ are determined classically by the method in Ref.~\cite{Babbush2018} based on Walker's alias method~\cite{Walker1974}, so that the final state of the circuit correctly approximate the desired one.
This part needs $m$ work qubits and $4L-4$ $T$ gates.
Then, we create the uniform superposition of $\mu$-qubit state, $\frac{1}{\sqrt{2^\mu}} \sum_{\sigma=0}^{2^\mu-1} \ket{\sigma}$ (the second top line of Fig.~\ref{fig:QROM_PREPARE}) and apply the inequality test circuit which compares the value of $\sigma$ and $\mr{keep}_l$ and flips the target bit (the most bottom line of Fig.~\ref{fig:QROM_PREPARE}) if $\sigma \geq \mr{keep}_l$.
The inequality test circuit uses $\mu$ work qubits and $4\mu-4$ $T$ gates~\cite{Berry2019qubitizationof}.
Finally, the controlled swap gates are applied to $\ket{l}$ on the original $m$ qubits and $\ket{\mr{alt}_l}$ on the ancillary $m$ qubits.
The conditional swap gates use $m$ Toffoli gates, which are decomposed into $7m$ $T$ gates.

In short, QROM-based construction of \prepare circuits requires  $N_a = 2\mu + m + 1$ ancillary qubits and $\max \{m+1, \mu \}$ work qubits, and 
\begin{align}
\begin{split}
    &(4m-4 + 2g_T(\epsilon_z)) + (4L-4) + (4\mu -4) + 7m\\ &= 4L + 4\mu + 11m - 12 + 2g_T(\epsilon_z)
\end{split}
\end{align}
$T$ gates.
This is $O(L) + O(\log(1/\epsilon))$ when we set $\mu = O(\log(1/\epsilon))$ and $\epsilon_z = \epsilon$.
Note that, although the choice of $\epsilon_z = \epsilon$ may seem too naive, $g_T(\epsilon_z)\approx 200$ even when $\epsilon_z=10^{-20}$ \cite{ross2014optimal}. Such change is negligible to the first term $4L$; for example, $4L>16000$ when $m=12$.

%%%%% -------------------- %%%%%
\section{Decomposition of real two-qubit gates \label{appsec:O4 decompose}}
The two-qubit gates $\mc{U}_s$ obtained by AQCE are real in our case, so it belongs to the orthogonal group $\textbf{O}(4)$. 
The following theorem is utilized to decompose such matrices into elementary gates~\cite{vatan2004optimal} in Step 2 of our proposal:
\begin{thm}
Let $M$ be
\begin{eqnarray} \label{eq:def of M}
    M = \frac{1}{\sqrt{2}} \begin{pmatrix}
        1 & i & 0 & 0 \\
        0 & 0 & i & 1 \\
        0 & 0 & i &-1 \\
        1 &-i & 0 & 0
        \end{pmatrix}.
\end{eqnarray}
Consider a matrix $U \in \mathbf{O}(4)$. When $\mathrm{det}(U)=1$, $M U M^\dagger$ is a tensor product of two unitary matrices in $\mathbf{SU}(2)$, i.e.,
\begin{eqnarray}
    M U M^\dagger \in \mathbf{SU}(2) \otimes \mathbf{SU}(2).
\end{eqnarray}
When $\mathrm{det}(U)=-1$, $MUM^\dagger U_{\mathrm{SWAP}}$, where $U_{\mathrm{SWAP}}$ is the swap gate, satisfies the same relation, i.e.,
\begin{eqnarray}
    M U M^\dagger U_{\mathrm{SWAP}}\in \mathbf{SU}(2) \otimes \mathbf{SU}(2).
\end{eqnarray}
\end{thm}

\begin{figure}
\includegraphics[width=\linewidth]{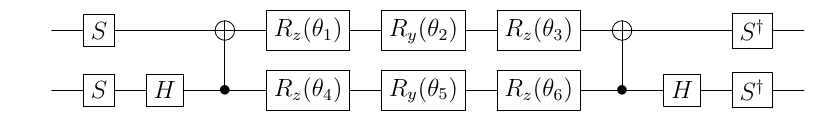}
  \caption{\label{fig:O4} Quantum circuit for decomposing a two-qubit gate in $\textbf{O}(4)$ with determinant +1.}
\end{figure}

\begin{figure}
\includegraphics[width=\linewidth]{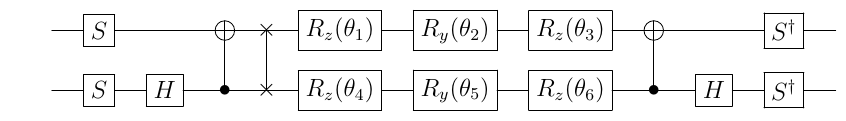}
 \caption{\label{fig:O4-1} Quantum circuit for decomposing a two-qubit gate in $\textbf{O}(4)$ with determinant -1.}
\end{figure}

This theorem enables us to decompose a two-qubit gate in $\mathbf{O}(4)$ into Clifford gates and six single-qubit rotations as depicted in Fig.~\ref{fig:O4} and \ref{fig:O4-1}.
This is because $M$ can be represented as a product of Hadamard (H), $S$, and CNOT gates as $M = \mathrm{CNOT}_{2,1} \cdot \mathrm{H}_2 \cdot S_{2} \cdot S_{1}$, and any $A \in \mathbf{SU}(2)$ can be written as the product of three single-qubit rotation $A=\rz(\theta) \ry(\theta') \rz(\theta'')$ for appropriate angles $\theta, \theta'$, and $\theta''$.

%%%%% -------------------- %%%%%
\section{Details of numerical experiments \label{appsec:details of numerics}}
\subsection{Geometries of molecules}
\begin{table*}[t] 
 \caption{Geometries of molecules. ``$(\mr{X}, (x,y,z))$" denotes three dimensional coordinates $x,y,z$ of atom X in units of \AA. 
 \label{tab: geometries}
 }
 \begin{tabular}{ccl}
 \hline \hline
 Molecule & Number of qubits & Geometry  \\ \hline
 \ce{LiH} & 12 & (Li, (0, 0, 0)), (H, (0, 0, 1.548)) \\
 \ce{H2O} & 14 & (O, (0, 0, 0.137)), (H, (0,0.769,-0.546)), (H, (0,-0.769,-0.546)) \\
 \ce{NH3} & 16 & (N, (0,0,0.149)), (H, (0,0.947,-0.348)), (H, (0.821,-0.474,-0.348)), (H, (-0.821,-0.474,-0.348)) \\
\ce{CH4} & 18 & (C, (0, 0, 0)), (H, (0.6276, 0.6276, 0.6276)), (H, (0.6276, -0.6276, -0.6276)), \\
 && (H, (-0.6276, 0.6276, -0.6276)), (H, (-0.6276, -0.6276, 0.6276)) \\
 \ce{CO} & 20 & (C, (0, 0, 0)), (O, (0, 0, 1.128)) \\
 \ce{H2S} & 22 & (S, (0, 0, 0.1030)), (H, (0, 0.9616, -0.8239)), (H, (0, -0.9616, -0.8239)) \\
 \ce{C2H2} & 24 & (C, (0, 0, 0.6013)), (C, (0, 0, -0.6013)), (H, (0, 0, 1.6644)), (H, (0, 0, -1.6644)) \\
 \ce{CO2} & 30 & (C, (0, 0, 0)), (O, (0, 0, 1.1879), (H, (0, 0, -1.1879)) \\
 \ce{C6H6} & 72 &
 (C, (0, 1.3868, 0)),
 (C, (1.2010, 0.6934, 0)),
 (C, (1.2010, -0.6934, 0)),
 (C, (0, -1.3868, 0)), \\ &&
 (C, (-1.2010, -0.6934, 0)),
 (C, (-1.2010, 0.6934, 0)),
 (H, (0, 2.4694, 0)),
 (H, (2.1386, 1.2347, 0)), \\ &&
 (H, (2.1386, -1.2347, 0)),
 (H, (0, -2.4694, 0)),
 (H, (-2.1386, -1.2347, 0)),
 (H, (-2.1386, 1.2347, 0))
\\ \hline \hline
 \end{tabular}
\end{table*}

The geometries of molecules other than the hydrogen chains are listed in Table~\ref{tab: geometries}.
These geometries are taken from CCCBDB database~\cite{NIST_CCCBDB} as the most stable structures at the level of Hartree-Fock/STO-3G.

\subsection{Generation of coefficients $c_l$}
The quantum chemistry Hamiltonian for molecules is in the form of
\begin{align}
\label{eq: original mol Ham}
\begin{split}
    H = & \sum_{\sigma=\uparrow, \downarrow} \sum_{p,q=0}^{N_{\mathrm{NO}}-1}  h_{pq}a^\dag_{p\sigma} a_{q\sigma}\\  
     & + \frac{1}{2} \sum_{\sigma,\tau=\uparrow, \downarrow} \sum_{p,q,r,s=0}^{N_{\mathrm{NO}}-1}  g_{pqrs}a^\dag_{p\sigma} a^\dag_{q\tau} a_{r\tau} a_{s\sigma}, 
\end{split}
\end{align}
where $N_{\mathrm{NO}}$ is the number of molecular orbitals,
$p,q,r,s$ are indices for the molecular orbitals, $\sigma=\uparrow, \downarrow$ represents the spin of electron, $h_{pq}$ and $g_{pqrs}$ are one- and two-electron integrals, and 
$a_{p\sigma}^\dag (a_{q\sigma})$ are fermion creation (annihilation) operators.

In our numerical experiments, the values of $h_{pq}$ and $V_{pqrs}$ are computed by PySCF \cite{Sun2018, Sun2020, Sun2015}.
From those values, we generate the coefficients $c_l$ by using Eq.~(129) of \cite{Mitarai2023perturbationtheory} under an assumption that all majorana operators in $H$ are mapped to distinct Pauli operators by some fermion-to-qubit mapping such as the Jordan-Wigner transformation \cite{Jordan1928}.

\bibliography{cite}

\end{document}